\documentclass[conference]{IEEEtran}
\IEEEoverridecommandlockouts

\usepackage{cite}
\usepackage{amsmath,amssymb,amsfonts}
\usepackage{algorithmic}
\usepackage{graphicx}
\usepackage{textcomp}
\usepackage{xcolor}
\def\BibTeX{{\rm B\kern-.05em{\sc i\kern-.025em b}\kern-.08em
    T\kern-.1667em\lower.7ex\hbox{E}\kern-.125emX}}
\begin{document}

\title{OAK - Onboarding with Actionable Knowledge}

\author{
\parbox[t]{0.30\textwidth}{
\centering
\textbf{Devènes Steve}\\
\fontsize{8pt}{9.5pt}\selectfont
Institute of Systems Engineering, HEI-VS\\
HES-SO University of Applied Sciences and Arts Western Switzerland\\
Switzerland\\
steve.devenes@hevs.ch
}
\hfill
\parbox[t]{0.30\textwidth}{
\centering
\textbf{Capallera Marine}\\
\fontsize{8pt}{9.5pt}\selectfont
HumanTech Institute, HEIA\\
HES-SO University of Applied Sciences and Arts Western Switzerland\\
Switzerland\\
marine.capallera@hefr.ch
}
\hfill
\parbox[t]{0.30\textwidth}{
\centering
\textbf{Cherix Robin}\\
\fontsize{8pt}{9.5pt}\selectfont
HumanTech Institute, HEIA\\
HES-SO University of Applied Sciences and Arts Western Switzerland\\
Switzerland\\
robin.cherix@hes-so.ch
}

\\[1.5em] 

\parbox[t]{0.30\textwidth}{
\centering
\textbf{Mugellini Elena}\\
\fontsize{8pt}{9.5pt}\selectfont
HumanTech Institute, HEIA\\
HES-SO University of Applied Sciences and Arts Western Switzerland\\
Switzerland\\
elena.mugellini@hefr.ch
}
\hfill
\parbox[t]{0.30\textwidth}{
\centering
\textbf{Abou Khaled Omar}\\
\fontsize{8pt}{9.5pt}\selectfont
HumanTech Institute, HEIA\\
HES-SO University of Applied Sciences and Arts Western Switzerland\\
Switzerland\\
omar.aboukhaled@hes-so.ch
}
\hfill
\parbox[t]{0.30\textwidth}{
\centering
\textbf{Carrino Francesco}\\
\fontsize{8pt}{9.5pt}\selectfont
Institute of Systems Engineering, HEI-VS\\
HES-SO University of Applied Sciences and Arts Western Switzerland\\
Switzerland\\
francesco.carrino@hevs.ch
}
}

\maketitle

\begin{abstract}
The loss of knowledge when skilled operators leave poses a critical issue for companies. This know-how is diverse and unstructured. We propose a novel method that combines knowledge graph embeddings and multi-modal interfaces to collect and retrieve expertise, making it actionable. Our approach supports decision-making on the shop floor. Additionally, we leverage LLMs to improve query understanding and provide adapted answers. As application case studies, we developed a proof-of-concept for quality control in high precision manufacturing.
\end{abstract}


\section{Introduction and Related Work}
``Employees recruited 3 months ago train employees recruited 3 weeks ago''. While somewhat caricatural, this statement collected during a workshop clearly illustrates a challenge that the industry, across various sectors (metallurgy, chemistry, pharmaceuticals, etc.), faces today: the loss of knowledge when experienced operators leave their jobs. This, combined with the current difficulty in recruiting a qualified workforce in certain sectors, poses a significant challenge.

Several studies conducted by the Federal Statistical Office (FSO), ETH Zurich, and Credit Suisse~\cite{CreditSuisse2019Monitor} confirm this issue and forecasts indicate that this phenomenon is expected to become increasingly significant. 

Practical experience of an operator and their specific knowledge (of a machine, a process, a methodology, etc.) consist of diverse and varied, unstructured information that is challenging to collect, standardize, search, and ultimately exploit. 

Knowledge bases (KB)~\cite{Pellissier2020YAGO} can be used to store, valorize, and exploit diverse, variable, and unstructured data with varying degrees of validity and truthfulness. Knowledge Graph Embedding (KGE) aims to integrate the components of a KB (entities and relations) into continuous vector spaces, simplifying their manipulation while preserving the inherent structure of the KB~\cite{Wang2017Knowledge, Bordes2011Learning, Zhang2018Learning}. The storage, processing and indexing of diverse data require management systems specifically designed for media (MM-DBMS) ~\cite{Thuraisingham2004MANAGING}, \cite{Ming1996Data}, and tools for managing and understanding user queries. 

To be ``actionable'', the stored information must be easily accessible by operators via the most suitable modality (e.g., text, speech) given the individual and environmental constraints (personal protective equipment, noise, lighting conditions, etc.). The use of multimodal interfaces, especially for human-robot collaboration \cite{Papanastasiou2019Towards}, \cite{Gunasekaran2019Multimodal} and Augmented Reality (AR) in industrial contexts has been explored in scientific literature \cite{Bottani2019Augmented}. Saidi et al. \cite{Saidi2022BL} studied integrating an educational AR application with existing maintenance software to easily retrieve maintenance data. However, there is no provision for recording instructions in their application, i.e., how to facilitate the storing of information coming from experienced operators and foster the creation of a company’s knowledge base. 

In this project, our aim is to combine novel open-source Large Language Models (LLM) architectures to build KGE, interrogate multimedia files, retrieve useful information, and present it to the user. Our approach is similar to retrieval-augmented generation (RAG) technics but based on KGE for the embeddings. In addition, we propose multimodal interfaces, AR and real-time object recognition to, a) allow experienced operators to record their know-how into the system and b) support unexperienced operators’ decision-making in real-time. Finally, we propose a Proof-Of-Concept (POC) around the use case of quality control in high-precision aluminum plate manufacturing, developed with the help of experienced operators.

\section{Method and implementation}
\begin{figure}
  \centering
  \includegraphics[width=1\linewidth]{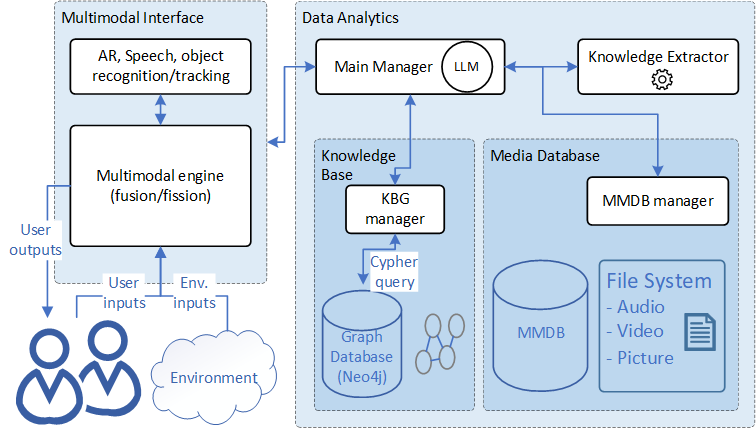}
  \caption{System Architecture}
  \label{fig:system_architecture}
\end{figure}

Fig. \ref{fig:system_architecture} presents the architecture of our solution. Our system is composed of two modules: \textbf{Data Analytics}, which handles the extraction, storage, and retrieval of knowledge; \textbf{Multimodal Interface}, which is the front-end part supporting several potential input modalities to record raw information from users and present back the results of their request. 

\subsection{OAK - Data analytics module }
The \textbf{Data Analytics} module is divided into four elements: an orchestrator (\textit{Main Manager}) and three connected modules assigned to handle specific tasks: the \textit{Media Database} is in charge of storing and indexing all raw media such as documents, pictures, audio and video files; the \textit{Knowledge Extractor} processes the raw media and retrieves information in triplet form; the \textit{Knowledge Base} implements a graph database (Neo4j \cite{Neo4j}, in our case) to store these triplets. If enabled, an LLM submodule within the \textit{Main Manager} interprets and translates users' queries and summarizing information for the user. 

\begin{figure}[!b]
  \centering
  \includegraphics[width=\linewidth]{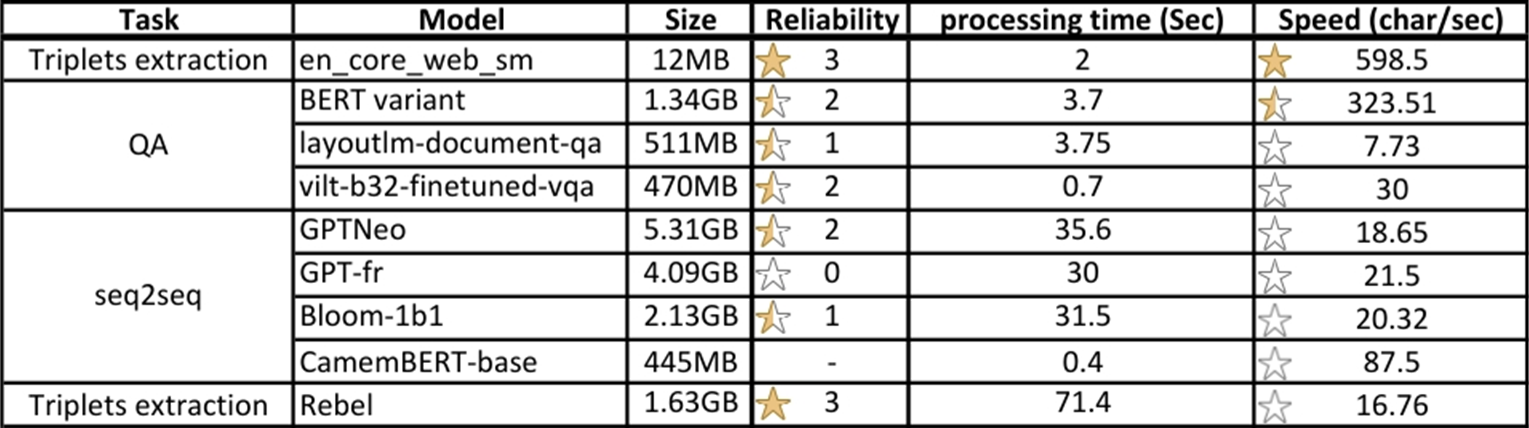}
  \caption{Model Comparison}
  \label{fig:model_comparison}
\end{figure}

For the \textbf{Data Analytics} module, we focused on two tasks: extracting knowledge from media and retrieving information using embeddings. We evaluated nine models, comparing their reliability, size, and normalized processing time (Fig. \ref{fig:model_comparison}). Reliability was subjectively scored from 0 to 3 (higher is better). To meet industrial data safety requirements, we only analyzed open-source models that could be installed on-premises.  

Finally, we retained \textit{Rebel} \cite{Huguet2021REBEL} because it got good scores and can directly generate triplets corresponding to the input text (which is not the case for \textit{en\_web\_core\_sm} \cite{spacy}). 

In a second step, we tested information retrieval via embedding models. These models encode data (images, text, etc.) in the form of vectors: similar vectors correspond to similar data, which can be used to calculate similarity scores. When applied to a graph database, embeddings can also be used to predict new relationships between entities, to classify entities or to reason about existing knowledge \cite{Wang2017Knowledge}.  

We tested four text embedding models. We limited our comparison to multilingual models able to understand French. The first three listed below are based on Sentence-BERT~\cite{reimers-2019-sentence-bert}: 
\begin{itemize}
    \item distiluse-base-multilingual-cased-v1
    \item paraphrase-multilingual-MiniLM-L12-v2
    \item paraphrase-multilingual-mpnet-base-v2 
    \item bloomz-560m-retriever \cite{DeBloomzRet}
\end{itemize}  

To test our approach, we created two ``toy'' benchmarks. The two benchmarks were quite different to evaluate the generalizability of this approach. The first used a graph database with 38 movies and 133 associated people (actors, producers, directors), provided as an example by Neo4j \cite{Neo4j}. For each movie, we created a textual context including the title, a short description, the release year, and a list of actors and directors. Then, we encoded these contexts with the models and posed generic questions about the movies. A similarity score for each question-context pair was calculated using cosine distance, ranging from -1 (opposite) to 1 (identical). By sorting contexts by their similarity scores, we could retrieve the most relevant information in the top results.

The second benchmark involved a dataset of 50 animal descriptions generated by ChatGPT v3.5. This test aimed to see if the system could identify an animal based on a visual description. Each description, consisting of short sentences, was used as context for the embedding models. We computed similarity scores against the embedded queries, similar to the movie dataset, to determine the most relevant matches.

Our qualitative assessment of the obtained results led us to choose \textit{distiluse-base-multilingual-cased-v1}. Its top-1 score was the highest alongside \textit{paraphrase-multilingual-mpnet-base-v2} but in terms of model’s size the former was half the latter. 

\subsection{OAK - Multimodal interface}
In an industrial setting, interfaces must be easy to use even with safety gear. They should show important information clearly without distracting operators. Safety gear limits the devices that can be used, so the interface needs to be simple and reduce errors from inaccurate movements. Since time is crucial, avoiding mistakes is important. The interface should help prevent unwanted actions, for instance, by spacing out buttons properly.
The \textbf{Multimodal Interface} module allows operators to interact via text, audio, or video as both input and output. The multimodal engine merges signals coming from different inputs (multimodal fusion) while providing to the users the information via the modality most suitable to the current context (multimodal fission). A transportable interface is desirable, as operators are likely to have to move around while consulting the knowledge base. It is also important to always leave at least one hand free for the operator. The device requires microphones and cameras to record the information needed to preserve the knowledge of experienced operators. AR-based solutions as well as speech or object recognition are also considered for hand-free interactions.  

\begin{figure}[!t]
  \centering
  \includegraphics[width=0.8\linewidth]{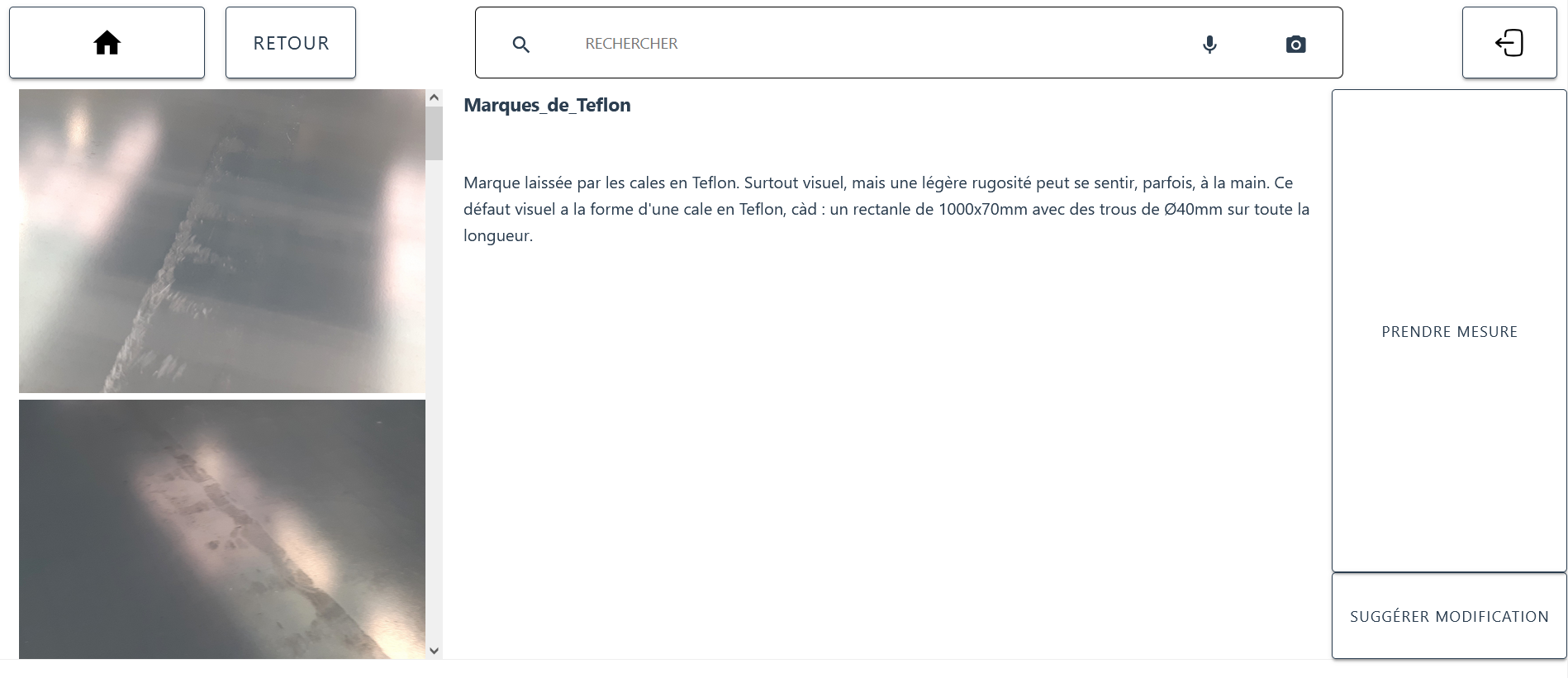}
  \caption{Example of defect description (French)}
  \label{fig:oakInterface}
\end{figure}

The main challenge of this project is helping novice operators quickly access useful information. We aim to improve access to relevant information using search and measurement tools with audio, images, and video (Fig. \ref{fig:oakInterface}). The interface should be familiar to the operators and involve them in its design to leverage their habits and reduce rejection. The interface should also involve the operators in the decision-making, information sharing process: they can send video commentaries on their actions (e.g., motivations or doubts), which can be used for reviews or to enrich the knowledge base. Additionally, a rating system among operators could help share and maintain valuable information. Once validated by experts, this rating could be used by the Data Analytics module to prioritize valuable or helpful information. 

\section{Test and preliminary results}
We implemented our solution in a real-world scenario: quality control of high-precision aluminum plate manufacturing. Implemented as a tablet application, our proof of concept (POC) supports operators of varying expertise in recognizing visual defects, assessing their significance, and making informed decisions (e.g., whether to scrap the entire piece). We created our knowledge base using the ``defects catalog,'' which is currently used in paper form on the shop floor to quickly identify defects, assess them, and guide operators in taking appropriate actions. The catalog includes short descriptions and sample images of defects.

\begin{figure}[!b]
  \centering
  \includegraphics[width=1.0\linewidth]{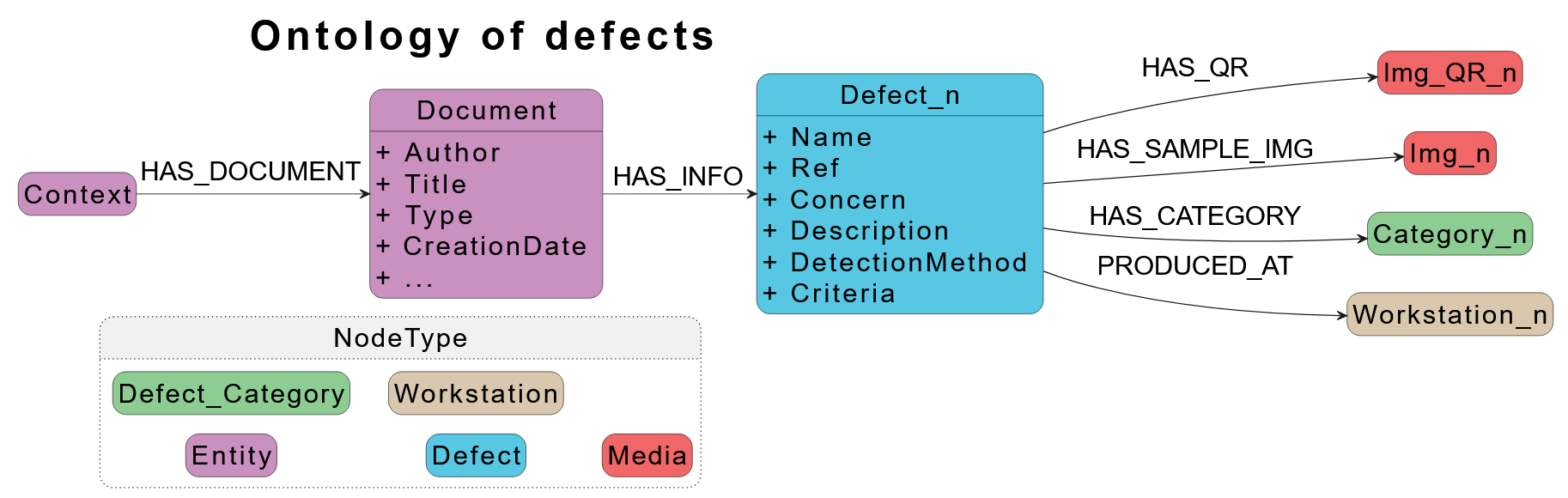}
  \caption{Ontology of defects}
  \label{fig:defect_ontology}
\end{figure}

Our KB stores defects, defect categories, machines associated with defect origins, and image IDs as nodes in a graph database, following the ontology shown in Fig. \ref{fig:defect_ontology}. Images are stored in the multimedia database (MMDB). Each image is assigned a unique identifier using a hash algorithm, preventing duplicates. Descriptions of defects are stored as embeddings linked to the corresponding defects, allowing operators to search the system by describing the defect they observe, either via text or voice. In addition, we integrated and trained a machine learning model into our POC for automatic visual defect recognition.

Alongside the defects’ catalog, other files representing experimented operators’ knowledge were provided. These files contain instructions on how to measure each defect, and the logical rules for determining the conformity of the plates according to the measurement values. 

\begin{figure}
  \centering
  \includegraphics[width=1\linewidth]{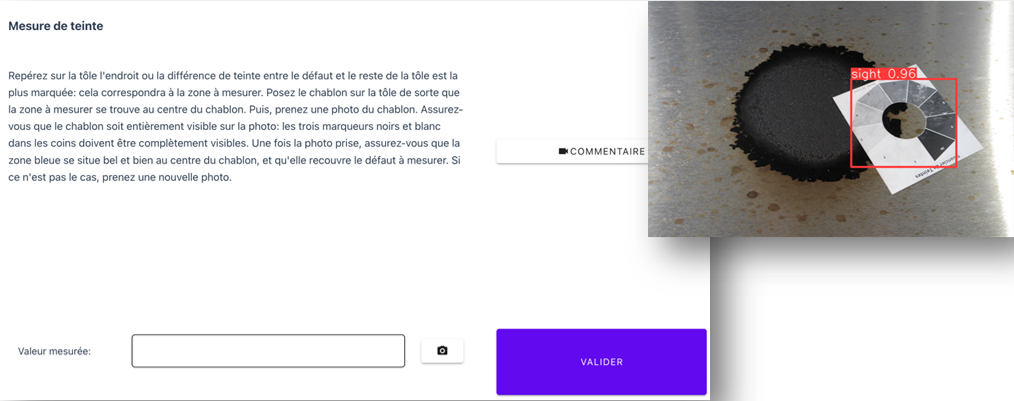}
  \caption{Guide (French) for stain defect assessment using AR and target tracking on a reference target. The interface allows adding video commentary to the measurement for validation and future training.}
  \label{fig:sight_detection}
\end{figure}

From an operator's point of view, the workflow is as follows: 
\begin{enumerate}
    \item Scan the ID of the product where a defect is spotted.
    \item Recognize the defect via textual, audio, or image-based description.
    \item Assist in the assessment of defect severity (see Fig. \ref{fig:sight_detection}).
    \item Log the measured value into the system, optionally providing an accompanying commentary.
    \item Receive a system suggestion concerning the conformity of the plate. The operator makes the final decision.
\end{enumerate}

\begin{figure}[!b]
  \centering
  \includegraphics[width=1\linewidth]{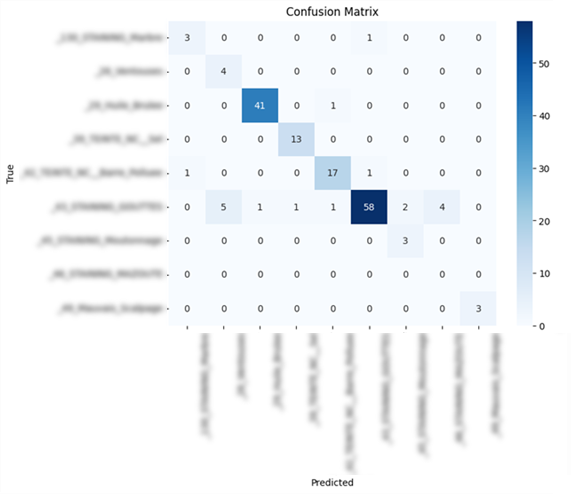}
  \caption{Confusion matrix. Defects' names are blurred for confidentiality reasons.}
  \label{fig:confusion_matrix}
\end{figure}

To evaluate our system, we used two distinct approaches. For the image-based search, we calculated the classifier score on a test set (Fig. \ref{fig:confusion_matrix}) focusing on 4 representative defects selected by our industrial partner. Given the small, unbalanced dataset used to fine-tune the Xception model~\cite{chollet2017xception}, the results are promising (weighted avg f1-score: 0.90).

For the audio and text-based search, a preliminary evaluation was conducted using the ``Top-n accuracy'' metric. We designed a questionnaire featuring five defects, two pictures per defect. For each defect, the questionnaire requested three different descriptions based solely on the presented images. Six individuals, completely novice in the field, completed the questionnaire, yielding a total of 88 descriptions (two left blank). Defect retrieval based on these descriptions resulted in the following: Top-1 accuracy: 0.19, Top-5 accuracy: 0.5, Top-10 accuracy: 0.67. The database contained a total of 28 distinct defects.
These mixed results can be explained by practical considerations: expert-provided text is often concise and technical, making direct correspondence with novice descriptions challenging.
Regarding interface evaluation, the User Experience (UEQ-S) generally averaged 1.69, with a \textit{pragmatic quality mean} of 1.42 and a \textit{hedonic quality mean} of 1.96. Pragmatically, operators found the application to be efficient (1.7), helpful and clear (1.5), and relatively simple (1.0). From a hedonic perspective, the solution was considered original (2.3), interesting (2.2), captivating and avant-garde (1.7). We used two items from the F-SUS questionnaire \cite{gronier2021psychometric} to gather qualitative feedback on the proposed solution. The scale ranged from 1 (``strongly disagree'') to 5 (``strongly agree'').
To the question ``I would frequently use this application,'' the mean response was (M = 4.17; std = 0.75).
To the question ``I felt confident using the application,'' the mean response was (M = 3.83; std = 0.75).
The highlighted advantages include ``time savings'', ``ease of processing'', and ``easier training for new colleagues''. The ``objectivity'' of measurement acquisition was also emphasized.

\section{Conclusion and future work}
We introduced a novel approach that uses knowledge graph embeddings and multimedia databases to store and retrieve the diverse and unstructured information representing operators’ know-how. This knowledge is made available to novice operators via a multimodal interface, which supports the novice operators’ decision-making via AR and real-time object tracking. We presented a POC tailored for a real-world scenario in the context of quality control in high-precision manufacturing. 

To further enhance the system, several avenues can be considered. Firstly, integrating more detailed and varied descriptions into the database, leveraging feedback from novice operators to refine existing descriptions, would be beneficial. Secondly, refining and training machine learning models with larger and more diverse datasets would improve defect recognition accuracy. Furthermore, incorporating video and other media types would enable more comprehensive and contextual defect analysis.
The user interface still requires testing and development iterations, prioritizing speed and, for the studied use case, a greater focus on image-based search.

\section*{Acknowledgment}

Left blank for blind review.

\bibliographystyle{IEEEtran}  
\bibliography{OAK}

\end{document}